# Feasibility Study of Microsecond Pulsed Microwave Ablation using a Minimally Invasive Antenna

Audrey L. Evans, *Student Member, IEEE*, James F. Sawicki, *Member, IEEE*, Hung Luyen, *Member, IEEE*, Yahya Mohtashami, *Member, IEEE*, Nader Behdad, *Fellow, IEEE*, and Susan C. Hagness, *Fellow, IEEE*

*Abstract*— In this study we established the feasibility of producing localized ablation zones using microsecond pulsed microwave ablation (MWA) as an alternative to conventional continuous wave (CW) MWA. We verified that a thin floating-sleeve dipole ablation probe can withstand pulsed power delivery with peak powers as high as 25 kW, with pulse widths on the order of 1 µs. We conducted MWA experiments in egg white using CW and pulsed modes of operation and found that ablation zones achieved via pulsed MWA are comparable in dimension to those created via CW MWA when the average power and procedure duration are equivalent. Finally, we performed pulsed MWA experiments in bovine liver and confirmed that pulsed MWA consistently produces large, localized ablation zones and temperatures that exceed 100°C. Establishing the feasibility of pulsed MWA opens the opportunity for developing a coupled MWA treatment and imaging system using pulsed MWA and microwave-induced thermoacoustic signals for real-time monitoring of MWA.

*Index Terms*—Microwave ablation, thermal therapy

## I. Introduction

MICROWAVE ablation (MWA) is a minimally invasive thermal therapy technique for treating cancer in human organs such as the lung, liver, kidney, and prostate [1]. An interstitial antenna radiates microwave energy into diseased tissue, rapidly heating the tissue via energy absorption and resulting in irreversible cell death of the target tissue. MWA therapy targets the diseased tissue in the vicinity of the active region of the antenna, while the surrounding healthy tissue remains undamaged [1].

Commercial MWA systems employ a continuous wave (CW) microwave source to deliver power to the tissue via the ablation antenna. A typical CW MWA system uses a sinusoidal signal that remains on for the duration of the procedure, corresponding to a 100% duty cycle. A lower duty cycle (~50%) can be advantageous for reducing cable heating or overcoming blood perfusion while maintaining the same average power [2], [3]. Typical average powers of CW MWA systems range from 10 to 100 W [4]. Most commercial MWA systems operate at either 915 MHz or 2.45 GHz, both of which fall within FCC industrial, scientific, and medical (ISM) bands. Laboratory MWA systems have also explored CW operation at higher frequencies to permit the design of smaller-footprint antennas [5], [6].

Alternatively, microwave energy can be delivered to tissue by means of a pulsed microwave system that has a high peak power level and a short (microsecond) pulse width such that the duty cycle is far below that used in quasi-CW operation. One benefit to using pulsed MWA is the byproduct of microwave-induced thermoacoustic signals. Tissue absorption of high-power microwave pulses induces the thermoelastic effect, resulting in the generation of thermoacoustic waves [7]. Typical peak power levels of pulsed microwave systems for the context of microwave-induced thermoacoustic imaging range from 1 to 300 kW, with pulse widths from 0.1 to 3 µs [7]–[11]. These conventional diagnostic, rather than therapeutic, systems use an external waveguide to excite the target; average power levels are low and the macro-scale temperature rise due to the microwave pulses is negligible. Thermoacoustic waves generated during pulsed MWA may enable ablation monitoring in real time, resulting in an integrated image-guided therapeutic system that uses the interstitial ablation antenna to simultaneously ablate tissue with microwave energy and generate acoustic signals that contain information about the spatial extent of the evolving ablation zone.

Lou and Xing (2010) reported a pulsed microwave thermal therapy experiment in which tissue was illuminated with pulsed microwave energy via an external waveguide [12]. A 6 GHz microwave source with a peak power of 300 kW, a pulse width

Manuscript submitted January 14, 2021. This material is based upon work supported by the National Science Foundation Graduate Research Fellowship under Grant No. DGE-1747503, a Vilas Associate Award from the University of Wisconsin-Madison, and the Philip D. Reed Professorship.

A. L. Evans, J. F. Sawicki, N. Behdad, and S. C. Hagness are with the Department of Electrical and Computer Engineering, University of Wisconsin-Madison, Madison, WI 53706 USA. (e-mail: alevans3@wisc.edu; jsawicki@wisc.edu, behdad@wisc.edu; susan.hagness@wisc.edu).

H. Luyen was with the Department of Electrical and Computer Engineering, University of Wisconsin-Madison, Madison, WI 53706 USA. He will be with the Department of Electrical Engineering, University of North Texas, Denton, Texas 76203 USA starting in January 2021. (e-mail: luyen@wisc.edu).

Y. Mohtashami was with the Department of Electrical and Computer Engineering, University of Wisconsin-Madison, Madison, WI 53706 USA. He is now with the University of California at Santa Barbara, Santa Barbara, CA, 93106 USA. (e-mail: y.mohtashami@ieee.org).

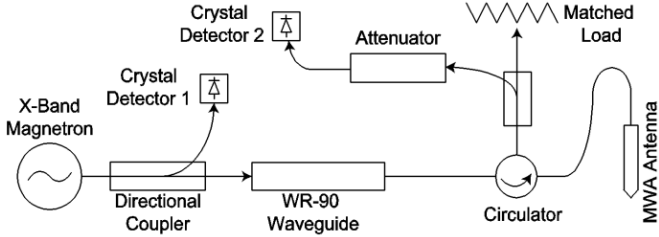

Fig. 1. The test setup used to perform pulsed MWA experiments.

of approximately 1 µs and a pulse repetition factor (PRF) of 15 Hz, 25 Hz, and 35 Hz delivered power to pork liver, with the corresponding average powers being 2.7 W, 4.5 W, and 6.3 W. The study reported that the tissue reached temperatures of up to 45°C. These reported temperatures are just below the threshold of cell necrosis at roughly 50°C [13].

In this paper we present a study on the feasibility of pulsed MWA using an interstitial antenna. There are two key feasibility considerations: the power handling capability of coaxial antennas when subjected to high power pulsed microwaves and the ability to achieve elevated temperatures necessary for generating ablation zones with clinically relevant dimensions. The power handling capability of the coaxial cable is limited by its dielectric strength. High-power pulses increase the dielectric breakdown risk of narrow-diameter coaxial antennas. Breakdown damages the dielectric material in an antenna, destroying the impedance match and creating a sudden increase in the reflected power.

We first demonstrated experimentally that a coaxial antenna designed for minimally invasive CW MWA can in fact withstand high peak powers during pulsed operation without breakdown. Then, we compared microsecond pulsed MWA to a conventional CW MWA setup and verified that pulsed MWA produces similarly sized ablation zone dimensions to CW MWA in egg white. We performed supplementary pulsed MWA experiments in bovine liver to further confirm that pulsed MWA can produce large ablation zone dimensions. Finally, we demonstrated the internal tissue temperatures during pulsed MWA reach temperatures that exceed 100°C within 60s, thereby inducing immediate cell necrosis.

## II. Methods

The experimental setup for our pulsed MWA experiments is shown in Fig. 1. Pulsed MWA experiments were conducted using an X-band pulsed magnetron with an operating frequency of 9.382 GHz, and a maximum pulse repetition rate of 909 Hz as the microwave source. The system was set up to operate with a peak power of 25 kW. At this peak power setting, the average output power for a pulse width of 1 µs and a duty cycle of 0.09% is 22.7 W. The peak power and pulse duration of this experimental system match typical parameters of existing microwave-induced thermoacoustic imaging systems [7]–[11] and the pulse repetition frequency was selected to achieve an average power level similar to typical CW MWA systems. The configuration included a directional coupler connected to a crystal detector for monitoring the incident pulse power, and a circulator and attenuator feeding into a second detector for monitoring reflected pulse power.

An X-band coaxial floating-sleeve dipole antenna was used as the MWA antenna. The dimensions and specifications of the antenna are reported in [6]. This antenna was selected for its robust CW MWA performance demonstrated previously in [6], [14]. It is one of many narrow-diameter minimally invasive designs that offers ease of fabrication. The maximum power handling capability of this antenna is calculated to be 78 kW based on the dielectric strength of the cable insulation material (PTFE), which is estimated to be 10 MVm$^{-1}$. Full-wave electromagnetic simulations of the antenna confirmed that the maximum electric field intensity throughout the entire antenna structure occurs along the inner conductor. The maximum operating temperature of the coaxial cable that the antenna is constructed from is 250°C [15]. We relied on calculations and simulation as a guide to predict the maximum power handling capability in lieu of stress testing the antenna at higher peak powers than what our available pulsed magnetron was capable of.

We conducted two sets of MWA experiments, first using egg white and subsequently using *ex vivo* bovine liver. Egg white was chosen for the first set of experiments because its translucency allows for real-time observation of the spatio-temporal evolution of the ablation zone for the duration of the experiment. The antenna reported in [6] is well matched in egg white and bovine liver at the operating frequency of 9.382 GHz.

### A. Experimental setup for pulsed and CW MWA in egg white

We designed the CW and pulsed MWA experiments in egg white such that the average power and duration was the same to permit a fair comparison of ablation zone dimensions. The power handling capability of the ablation antenna was also evaluated during the pulsed MWA experiment. The pulsed ablation experiment was conducted for a duration of 10 minutes using a peak power of 25 kW, pulse duration of 0.9 µs, pulse repetition rate of 900 Hz, and average power of 20 W. Photographs of the ablation zone in the translucent egg white were taken at 0, 5, and 10 minutes. The CW ablation experiment was conducted using a CW microwave generator at 9.382 GHz and power of 20 W for a duration of 10 minutes. The incident and reflected power from the antenna, measured at crystal detector 1 and crystal detector 2 respectively, as shown in Fig. 1, were recorded. The $|S_{11}|$ of the antenna was measured before and after the pulsed MWA experiment.

### B. Experimental setup for pulsed MWA in bovine liver

A series of pulsed MWA experiments were conducted in *ex vivo* bovine liver with a peak power of 25kW and pulse duration of 1 µs. Three different average powers of 22.7 W, 17.0 W, and 11.4 W were achieved by altering the pulse repetition rate. A Neoptix fiber-optic temperature sensor was positioned with tape to the outer surface of the active region of the antenna (longitudinally 6 mm from the tip of the antenna) to monitor temperature during each experiment as a function of time. After each ablation experiment in bovine liver, we dissected the tissue sample along the plane of the insertion path of the antenna to record ablation zone dimensions. The length of the ablation zone is the dimension along the insertion path of the antenna and the width was recorded as the widest part of the ablation zone perpendicular to the antenna insertion. The dimensions of each ablation zone were recorded and photographed. A summary of the experimental trials for bovine liver is presented in Table 1. The experimental trials were designed to cover both fixed power levels with varying durations as well as varying power levels to permit a variety of comparisons. The ablation zone dimensions and internal temperature were recorded for each experimental trial.

TABLE 1
SUMMARY OF EXPERIMENTAL TRIALS FOR PULSED MWA OF BOVINE LIVER

| Pulse repetition rate | Peak pulse power | Average power | Duration | Number of experiments |
|---|---|---|---|---|
| 909 Hz | 25 kW | 22.7 W | 2 min | 6 |
| 909 Hz | 25 kW | 22.7 W | 6 min | 14 |
| 682 Hz | 25 kW | 17.0 W | 8 min | 6 |
| 454 Hz | 25 kW | 11.4 W | 12 min | 6 |

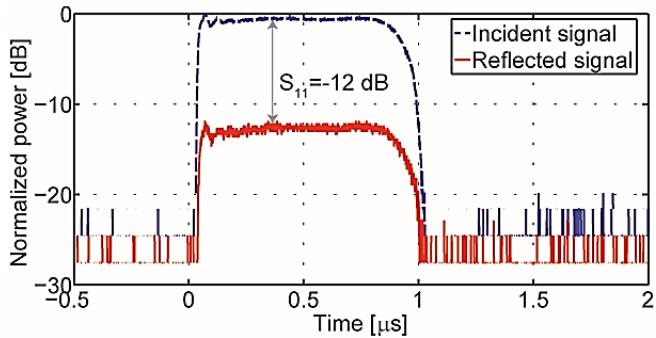

Fig. 2. The incident and reflected pulse power of the ablation antenna during a typical 1 µs pulse used in pulsed MWA experiments.

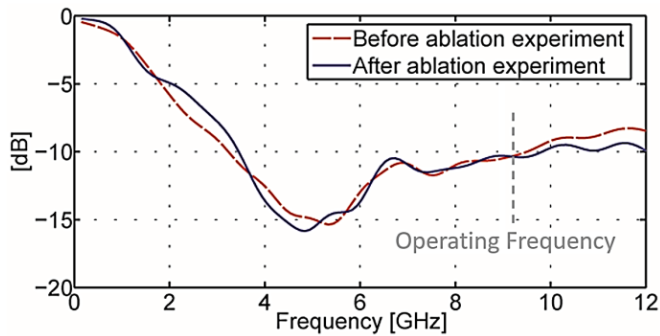

Fig. 3. The measured wideband $|S_{11}|$ before and after a 10-minute pulsed MWA procedure with peak power of 25 kW, pulse duration of 1 µs, and a duty cycle of 0.09%.

### III. RESULTS

#### A. Power handling capability of coaxial ablation antenna

Our first step in verifying experimentally that the coaxial ablation antenna does not breakdown during high-peak power pulses was to confirm that the reflected power from the antenna remained constant for the duration of the 1 µs pulse [16]. Fig. 2. shows a typical incident and reflected pulse measured during a pulsed MWA experiment. The magnitudes of the pulses are normalized to the peak power level of 25 kW. The magnitude of the reflected pulse is approximately 12 dB below the incident pulse, indicating a good impedance match. The reflected power was stable throughout the 1 µs pulse, confirming that the antenna withstood pulses with peak power levels of 25 kW without breakdown. Fig. 3 shows the measured $|S_{11}|$ of the ablation antenna before and after a 10-minute pulsed MWA experiment using 25 kW peak power and a pulse repetition rate of 909 Hz. Despite the change in tissue dielectric properties before and after ablation, the reflection coefficient of the antenna demonstrates only a small variation of the antenna impedance before versus after ablation. The primary reason for this small change is that the antenna's response is dominated by the materials closest to the metal, which is the Teflon insulation, rather than the tissue [17]. At the operating frequency, the antenna is well matched at -11 dB before and after ablation, indicating that the antenna did not undergo degradation or breakdown during the pulsed MWA experiment. Furthermore, antennas used repeatedly (20+) for pulsed MWA experiments maintain an acceptable $|S_{11}|$, indicating that they have not undergone breakdown. This demonstration establishes the suitability of this MWA antenna system for high-peak-power applications.

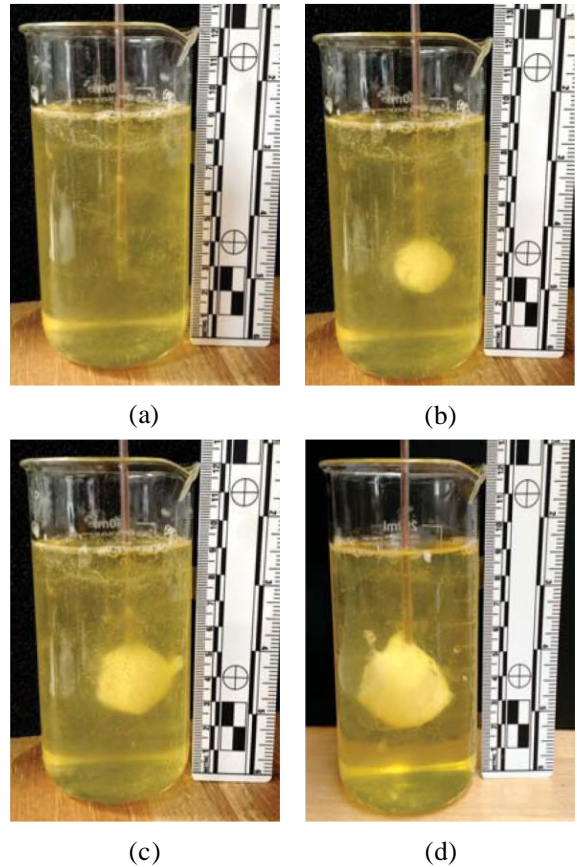

Fig. 4. Ablation zones creating in egg white using pulsed MWA for a duration of (a) 0 minutes, (b) 5 minutes, and (c) 10 minutes compared (d) to a 10-minute CW MWA experiment. The average power in both the pulsed and CW experiments was fixed at 20 W.

TABLE 2
DIMENSIONS OF PULSED ABLATION ZONES IN BOVINE LIVER

| Average power [W] | Time [min] | Length [mm] | | | Width [mm] | | |
|---|---|---|---|---|---|---|---|
| | | Mean | Max value | Min value | Mean | Max value | Min value |
| 22.7 | 2 | 15.7 | 19 | 12 | 12.7 | 14 | 11 |
| 22.7 | 6 | 26.7 | 31 | 25 | 19.3 | 23 | 17 |
| 17.0 | 8 | 21.5 | 23 | 20 | 18.7 | 20 | 13 |
| 11.4 | 12 | 19 | 22 | 18 | 16 | 19 | 14 |

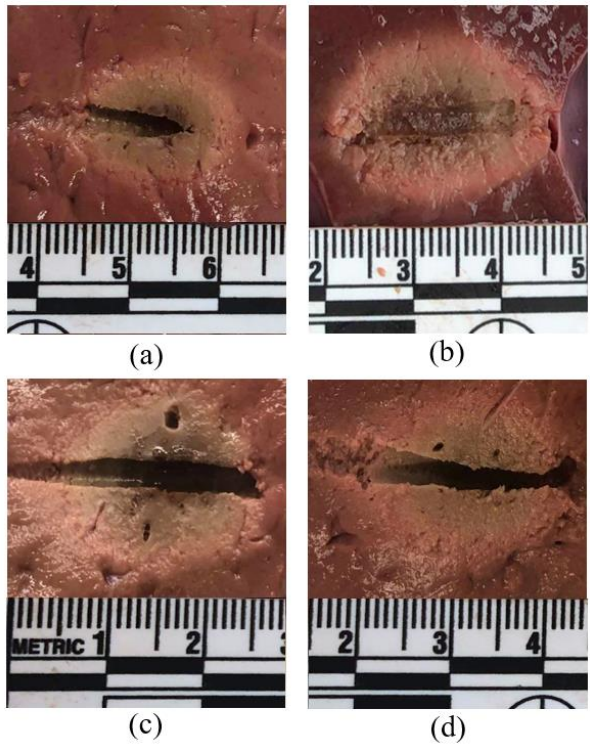

Fig. 5. Representative pulsed MWA zones in bovine liver using an average power and duration of (a) 22.7 W, 2 min, (b) 22.7 W, 6 min, (c) 17.0 W, 8 min, and (d) 11.4 W, 12 min.

*B. Feasibility of pulsed MWA in egg white*

Photographs of pulsed MWA experiments in egg white at 0, 5, and 10 minutes are shown in Fig. 4 (a-c). Fig. 4 (d) shows the result of a CW MWA experiment at the same average power after 10 minutes. The roughly spherical ablation zones were observed to be comparable in size at the same time point for both pulsed MWA and CW MWA. We estimated that the diameters of the ablated egg white for the pulsed and CW cases to be 2.7 cm and 3.0 cm, respectively. A 10% ablation diameter variation is common between CW MWA experiments due to subtle variations between tissues under test and is not considered to be significant [4]. These experiments demonstrate that at an average power of 20 W, pulsed MWA can produce similar sized ablation zones using a peak power of 25 kW, a pulse repetition frequency of 909 Hz, and a duty cycle of 0.09% compared to a CW experiment using a duty cycle of 100%.

*C. Feasibility of pulsed MWA in bovine liver*

Representative photographs of pulsed MWA experiments in bovine liver are presented in Fig. 5. Table 2 summarizes the average, maximum, and minimum ablation zone dimensions (length and width of the bisection plane) for all experiment groups. The experiments conducted operated at 22.7 W for durations of 2 minutes and 6 minutes illustrate that ablation zone increases in size over time, as shown in Figs. 5(a) and 5(b). Fig. 6 compares the internal ablation temperature as a function of time for the 2-, 8-, and 12-minute pulsed MWA trials. For all three average power levels the antenna takes less than 60 s to reach temperatures well above the threshold of cell necrosis. At this highest average power level, the internal temperature plateaus at approximately 100°C after ~120 s. As the average power decreases, the temperature rise is more gradual over time. The rapid heating with pulsed MWA further validates the feasibility of producing centimeter-sized ablation zones.

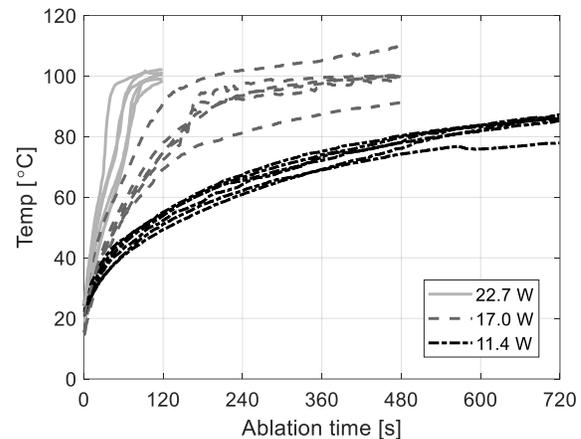

Fig. 6. Temperature at the center of the active region of the antenna during pulsed MWA experiment trials in bovine liver for three different average powers.

## IV. Conclusion

We demonstrated the feasibility of microsecond pulsed MWA, an alternative energy delivery approach for producing comparable ablation zones to conventional CW MWA systems. The potential risk for antenna breakdown at high peak powers was addressed, and we found that the narrow-diameter floating sleeve coaxial dipole antenna designed for MWA does not experience degradation when subjected to microsecond pulses with a peak power of 25 kW and temperatures as high as 110°C. Ablation zone dimensions for pulsed and CW MWA experiments in egg white using the same average power and duration were comparable. Our pulsed MWA experiments in bovine liver further established the consistency of large and generally spherical ablation zones. This demonstration establishes the foundation for future work coupling ablation and monitoring-signal excitation using thermoacoustic signals that are produced from pulsed microwave energy.

## Acknowledgment

The authors would like to thank Dr. Tyler Rowe for his technical assistance in setting up the pulsed magnetron used in the MWA experiments reported in this study.